# Relaxation under Supercooling of the Disordered Vortex State in Doped CeRu$_2$ Alloy

**M. K. Chattopadhyay, K. J. S. Sokhey, S. B. Roy, and P. Chaddah**
Low Temperature Physics Laboratory, Centre for Advanced Technology, Indore 452 013, India

*Abstract*

*Relaxation in magnetisation of metastable states is reported. The states are prepared by supercooling a Ce(Ru, 5% Nd)$_2$ alloy sample to different extents through similar paths in the H-T space, and then by applying a field jerk of ±20 Oe. Based on standard theory, a possible explanation is put forward.*

## INTRODUCTION

The onset of peak-effect (PE) in the J$_C$ of CeRu$_2$ is accompanied by a first-order phase transition (FOPT) in the vortex state [1-3]. Supercooling has been observed across this transition both by decreasing field H and by decreasing temperature T, with the latter procedure allowing much deeper supercooling [3]. Standard phenomenology of FOPT has been used to study metastability under different paths in H-T space, and under different extents of supercooling [4-7]. Measurements [3] on CeRu$_2$ have supported the prediction [4] that supercooling by reducing H reduces the effective free-energy-barrier seen by the metastable state. Measurements on V$_3$Si [8] have supported the expectation [5] that a more deeply supercooled state is unstable to a smaller fluctuation. In this paper we study relaxation in metastable states that have been supercooled to different extents, but following the same path in H-T space. As per the expectation [5] that more deeply supercooled states see a smaller free-energy-barrier, the relaxation rates should be dependent on the extent of supercooling.

## EXPERIMENTAL

In these measurements we have used a CeRu$_2$ sample with 5% Nd substitution on Ce site, as has been used in our earlier measurements [1,2,9]. This sample has been used as a representative of the CeRu$_2$ family in many of our earlier measurements as well, since in the pure sample the PE region encompasses both para and diamagnetic regimes over a small change in applied magnetic field; and SQUID measurements involving such a small variation in applied field can lead to uncertainties in the crossover regime. In the presently used sample, the PE region (which possesses all the characteristics observed for pure $CeRu_2$ [1,2]) is confined to the paramagnetic regime alone. DC magnetization measurements were performed using a Quantum Design MPMS-5 SQUID magnetometer with a scan length of 2 cm. For relaxation studies in the disordered state under various extents of supercooling the field H was applied at T = 10K, well above $T_C$ = 6.8K, and the temperature was lowered monotonically to 5K, thus following path A indicated in fig 1. For H values, lying within the schematic $H_C$ and $H^*$ lines in fig 1, this prepares the initial supercooled state, with lower values of H giving states that are more deeply supercooled. We wait for 1800 seconds after the temperature is stable at 5K, and then apply a field change h of magnitude 20 Oe and of either sign. This is greater than the field for full-penetration at all values of H studied [9], and thus the critical state is set up. Relaxation of magnetization was measured in this critical state, for various values of H, and for both signs of h at each H.

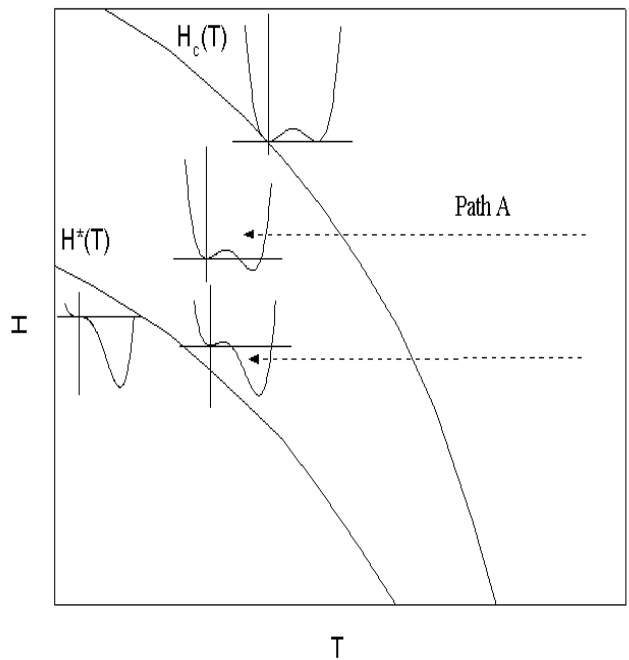

**Fig 1** Schematic showing the extent of supercooling.

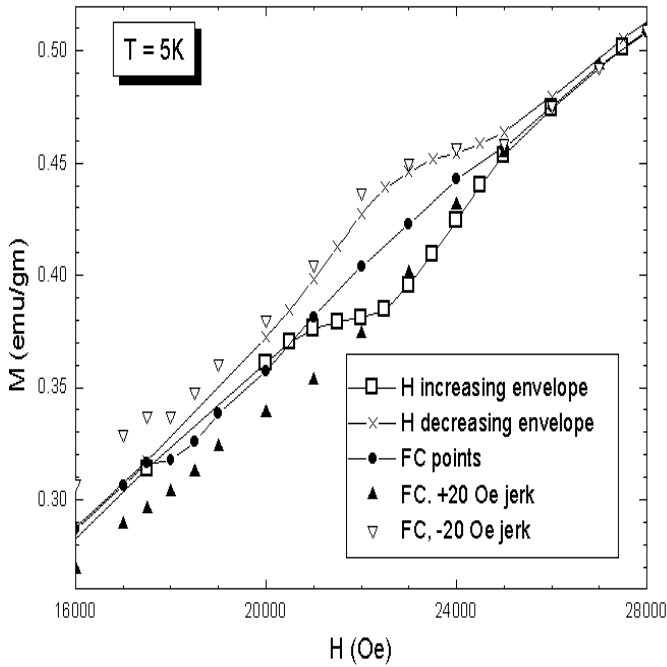

**Fig 2** Variation of magnetisation with applied field in the Peak Effect regime.

## RESULTS & DISCUSSION

We show in figure 2 the isothermal M-H scan at 5K, highlighting the PE regime. As has been recorded earlier [9], $J_C$ shows a peak at 22.5 kOe. We also show the values of M in the critical state created after field cooling, and the difference in the M-values for the two signs of h provides a measure of $J_C$ in the supercooled disordered state. As is apparent, $J_C$ following field cooling is enhanced in comparison with the $J_C$ obtained in the isothermal scan. Steingart et al [10] had observed the same feature in transport measurements on niobium and had defined an 'enhancement factor'. The enhancement factor for the present sample, as can be extracted from fig 2, peaks at 17.5 kOe i.e. at a field much lower than that (22.5 kOe) at which the isothermally determined $J_C$ peaks. This is consistent with the data reported earlier in fig 5 of ref [9], and we shall come back to this.

We now concentrate on our data on the relaxation of M(t). The data can be fit to the usually observed logarithmic decay with time. We plot in fig 3 the logarithmic decay rates dM/d (ln t) for various values of H between 16 kOe and 28 kOe, encompassing the field range in which the PE is seen in the isothermal scan. The square symbols show the decay rate for h = +20 Oe, and the circular symbols show the rate for h = -20 Oe. We make the following observations from this data:

1. The relaxation rate drops sharply for H at 22 kOe and above, while $J_C$ peaks at 22.5 kOe (see fig 5 of ref [9]). As noted earlier, the enhancement factor becomes negligible at 23 kOe [9], indicating that field-cooled states at this and higher fields are the stable equilibrium states.
2. The relaxation rate is large below 22 kOe, and peaks around 17.5 kOe, whereas the PE in the isothermal scan is initiated only above 20 kOe. This is also the field regime in which the enhancement factor is large [9]. This indicates clearly that field cooling in this field regime results in metastable supercooled states.

3. The relaxation rate is rising as H decreases from 21 kOe to 18 kOe.
4. The relaxation rate drops sharply as H is lowered to 17 kOe and below.

We now provide a plausible explanation for these observations within the standard phenomenology of FOPT. We conjecture based on observation 1 that at 5K $H_C$ is about 22 kOe, so that the field-cooled state for H > 22 kOe is the stable disordered state. We also conjecture based on observation 4 that the value of $H^*$ at 5K is about 17 kOe, so that the field-cooled state for H < 17 kOe is the stable ordered state. With these two values for $H_C$ and $H^*$, we can explain the observations 2 and 3. For 17 kOe < H < 22 kOe, the field-cooled state is the metastable (supercooled) disordered state with its critical state having a large relaxation rate. The free-energy-barrier decreases as H is lowered because one is monotonically moving from $H_C$ (where the barrier is large) to $H^*$ (where the barrier is zero) [4], and this explains why the relaxation rate rises as H is lowered in this range. This also explains the result of ref [9], mentioned above, that the enhancement factor is larger at 17.5 kOe than at 22.5 kOe.

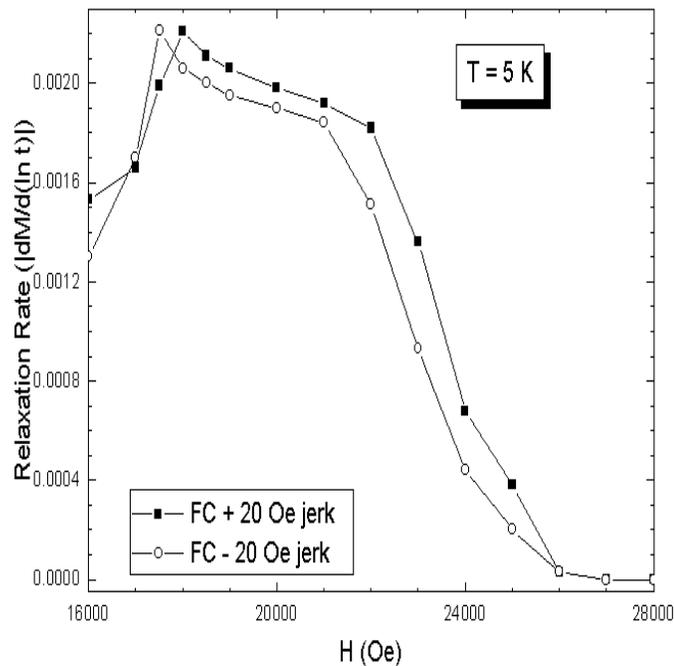

**Fig 3** Variation of relaxation rate with applied magnetic field.

While we have provided a simple explanation for the observations, we shall end with the caution that in samples with non-zero $J_C$ the applied field is not the same as the local field everywhere, and this results in broadened bands for $H_C$ and $H^*$ [6,7]. We thus have metastable disordered regions, within the sample, persisting at fields below that where other regions have transformed to the stable ordered state.